\documentclass[aps, pra, a4paper, twocolumn, noshowpacs, superscriptaddress, floatfix]{revtex4}
\usepackage{graphicx}
\usepackage{amsmath}
\usepackage{amssymb}

\begin{document}
    \title{The intrinsic relations of quantum resources in multiparticle systems}
  
    \author{Wen-Yang Sun} 
        \affiliation{School of Physics \& Material Science, Anhui University, Hefei 230601,  China}
   
    \author{Dong Wang}
       \affiliation{School of Physics \& Material Science, Anhui University, Hefei 230601, China}
       \affiliation{CAS Key Laboratory of Quantum Information, University of Science and Technology of China, Hefei 230026, China}
       
     \author{Bao-Long Fang} \
       \affiliation{Department of Mathematics \& Physics, Hefei University, Hefei 230601, China}
       
    \author{Zhi-Yong Ding}
	    \affiliation{School of Physics \& Material Science, Anhui University, Hefei 230601, China}
	    \affiliation{CAS Key Laboratory of Quantum Information, University of Science and Technology of China, Hefei 230026, China}
	        
    \author{Huan Yang}
	    \affiliation{School of Physics \& Material Science, Anhui University, Hefei 230601,  China}
	    
	\author{Fei Ming}
		\affiliation{School of Physics \& Material Science, Anhui University, Hefei 230601, China}
	
	\author{Liu Ye}  
		\email[Corresponding author: ]{yeliu@ahu.edu.cn (L. Ye)}
		\affiliation{School of Physics \& Material Science, Anhui University, Hefei 230601,  China}
 
	\begin{abstract}
	\textbf{Abstract:} Quantum resources play crucial roles for displaying superiority in many quantum communication and computation tasks. To reveal the intrinsic relations hidden in these quantum resources, many efforts have been made in recent years. In this work, we investigate the correlations of the tripartite \textit{W}-type states based on bipartite quantum resources. The interrelations among the degree of coherence, concurrence, Bell nonlocality and purity are presented. Considering Bell nonlocal and Bell local (satisfied the Clauser-Horne-Shimony-Holt inequality) states for the two-qubit subsystems derived from the tripartite \textit{W}-type states, we find exact lower and upper boundaries of the degree of coherence versus concurrence. Interestingly, exact relation  among the degree of coherence, concurrence and purity is obtained. Moreover, coherence is also closely related to entanglement in two specific scenarios: the tripartite \textit{W}-type state under decoherence and a practical system for a renormalized spin-1/2 chain.
		
	
	\end{abstract}
    \maketitle
      
   	\section{Introduction} \label{sec1}
	    Multiparticle quantum systems can show unique forms of quantum nonlocal correlations (QNCs) \cite{w01,w02,w03,w04,w05}. QNCs and coherence are fundamental quantum resources (QRs) in many novel quantum-information tasks that cannot be achieved by classical resources. For bipartite systems shared by two parties, Alice and Bob, many efforts have been devoted to a deeper insight of nonlocal correlations, mainly in consideration of three types of QNCs: entanglement \cite{w02,w03}, Bell nonlocality \cite{w04,w05} and Einstein-Podolsky-Rosen (EPR) steering \cite{w06, w07, w08, w09}. 
	    
	   Entanglement, serve as one of the most generally used QRs, is defined as the inseparability of quantum states and can be considered as an algebraic concept \cite{w03}. Besides, another one quantum resource  can be discovered by violating some Bell-type inequalities, and is termed as Bell nonlocality \cite{w10,w11,w12,w13,w14,w15,w16,w17}. One of these Bell-type inequalities, the Clauser-Horne-Shimony-Holt (CHSH) inequality \cite{w03,w04,w05} has been often considered to measure the QNCs presented between spatially separated two parties that are entangled, and thereby it can violate the bound originated by the inequality \cite{w16}. It is noteworthy that, Bell-type inequalities can be violated only if their states are entangled. However, there are entangled states that can still exhibit QNCs which cannot violate any Bell-type inequality for any possible local measurements proposed by Werner \cite{w18} in 1989. Then, the sufficient and necessary conditions for arbitrary bipartite states to be Bell nonlocal states were derived \cite{w19} in 1995. The last one is called as EPR steering \cite{w06, w07}, which was introduced by Schr\"{o}dinger \cite{w20} to analyze the EPR-paradox in 1935. Conceptually, EPR steering is able to describe that an observer can instantaneously affect a remote system by utilizing local measurements. In addition, EPR steering has been viewed as an intermediate typr of quantum resource \cite{w21, w22, w23} between entanglement and Bell nonlocality in modern quantum-information theory. Except for their fundamental importance, these QRs have had many practical applications in quantum-information-science ranging from quantum-key-distribution \cite{w10, w22}, quantum random-number generation \cite{w11, w12}, and communication complexity \cite{w24}.
	    
	    On the other hand, coherence, as one of the most widely applied QRs, plays a key role in the range of biological, chemical and physical phenomena \cite{w25}, such as, quantum metrology \cite{w26}, low-temperature thermodynamics \cite{w27,w28,w29}, solid-state physics \cite{w30} and also explains the violation of the CHSH inequalities. Meanwhile, coherence is also an important concept used to depict the interference capability of interacting fields in quantum optics research \cite{w31,w32,w33}. Additionally, the knowledge of the internal distribution of coherence between subsystems and their correlations becomes essential for predicting the coherent evolution in the researched quantum system \cite{w34,w35}.  
	    
	    Even though these QRs are playing different roles in different quantum-information tasks, the internal relations behind the QRs might be the same one. Recently many researchers have made progress for this target \cite{w35, w36, w37, w38, w39, w40, w41, w42}. For example, Kalaga \textit{et al.} \cite{w38} have studied the relations between EPR steering and coherence in the families of the tripartite entangled states, and they disclose some mutual relations among the EPR steering, entanglement, and coherence. Beside, another result is generalized to quantum discord (QD) and promoted to multipartite systems in Ref. \cite{w39}, which it is indicated that QD generated by multipartite incoherent operations is restricted by coherence expended in its subsystems. All the works above are trying to connect part of these QRs. Notwithstanding, coherence, entanglement and QD are qualitatively unified in interferometric framework \cite{w42}, quantitative relations of these QRs is still an open question.
	    
	    In this work, we will concentrate on finding mutual relations of these QRs of bipartite states. Particularly, we are interested in the quantitative relations among Bell nonlocality, entanglement, coherence and purity based on bipartite subsystems in a general multiparticle system (a tripartite \textit{W}-type states). Here we severally consider Bell nonlocal and Bell local (satisfied the CHSH inequality) bipartite mixed states in the tripartite \textit{W}-type states, and analyze the mutual relations between the degree of coherence and concurrence. Interestingly, quantitative relation among the degree of coherence, concurrence and purity is revealed. Additionally, we consider two specific  scenarios, one scenario is a  tripartite \textit{W}-type state under decoherence channel, we research how the phase flip channel influences the quantitative relations among these QRs. And the other one is for a practical system of a renormalized spin-1/2 Heisenberg \textit{XXZ} model, in this spin system, some mutual relations between QRs and purity are attained.
	    
	    The remainder of this article is organized as follows. Preliminary definitions and notations are introduced in Sec. \ref{sec2}. Then, we investigate the internal relations among these QRs in the case of the qubit-qubit subsystem, and derive some exact boundary conditions with respect to Bell nonlocal and Bell local states in Sec. \ref{sec3}. In Sec. \ref{sec4}, we discuss two specific scenarios for the tripartite \textit{W}-type states (under a decoherence channel and a renormalized spin-1/2 Heisenberg \textit{XXZ} model). In final, we end up our paper with a brief conclusion.
	   	    	
   	\section{Preliminaries} \label{sec2}
        We consider a $2\times 2$ dimensional quantum state ${\rho _{AB}}$, composed of subsystems \textit{A} and \textit{B}. Each subsystem is characterized by the corresponding density matrix,  ${\rho _A}$ and  ${\rho _B}$. The degree of first-order coherence of each subsystem \textit{A} and \textit{B} is a better methodology for quantifying this coherence, and it can be given by \cite{w33, w34}
    	\begin{equation}\label{E01}  
    		{D_\kappa } = \sqrt {2{\rm{Tr}}(\rho _\kappa ^2) - 1} ,{\rm{ }}\kappa  = A,{\rm{ }}B.
    	\end{equation} 
    	Then, one can introduce a measure of coherence named the degree of coherence (DC) for both subsystems \textit{A} and \textit{B} when they are considered independently \cite{w35,w38}
    	\begin{equation}\label{E02}  
    	D_{AB}^2 = \frac{{D_A^2 + D_B^2}}{2},
    	\end{equation} 
    	when both subsystems are coherent, we have  $D_{AB}^2 = 1$, while only if both subsystems show no coherence, $D_{AB}^2 =0$. Besides, we use a popular entanglement measure, Wootters' concurrence \cite{w43,w44}. The Wootters' concurrence can be defined by
    	\begin{equation}\label{E03}  
    	{C_{AB}} = \max \left\{ {0,\sqrt {{\lambda _1}}  - \sqrt {{\lambda _2}}  - \sqrt {{\lambda _3}}  - \sqrt {{\lambda _4}} } \right\},
    	\end{equation}
    	 where  ${\lambda _i}$ are the decreasing-order eigenvalues of the matrix $R = {\rho _{AB}}({\sigma _y} \otimes {\sigma _y})\rho _{AB}^*({\sigma _y} \otimes {\sigma _y})$. This measure is a monotonic and convex function of the entanglement of formation \cite{w45}.    	 
    	     
    	 Next, it is well known that the violation of Bell's inequality in the form given by Clauser, Horne, Shimony, and Holt in two-qubit systems means that the entanglement for the bipartite appears, but not vice versa, \textit{i.e.}, there are entangled states which do not violate the CHSH inequality. The CHSH inequality can be expressed as \cite{w03,w46}
    	 	\begin{equation}\label{E04}  
            \left| {\left\langle {{B_{CHSH}}} \right\rangle } \right| = \left| {{\rm{Tr}}\left( {{\rho _{AB}}{B_{CHSH}}} \right)} \right| \le 2,
    	   \end{equation} 
    	 depending upon the CHSH operator ${B_{CHSH}} = a \cdot \sigma  \otimes (b + b') \cdot \sigma  + a' \cdot \sigma  \otimes (b - b') \cdot \sigma $  where $a,{\rm{ }}a'$  and $b,{\rm{ }}b'$  are unit vectors describing the measurements on sides \textit{A} and \textit{B}, respectively. In terms of the Horodecki theorem \cite{w19}, the maximum expected value of the CHSH operator for the quantum state ${\rho _{AB}}$  can be given by
    	  \begin{equation}\label{E05}  
    	  \mathop {\max }\limits_{{B_{CHSH}}} \left| {{\rm{Tr}}\left( {{\rho _{AB}}{B_{CHSH}}} \right)} \right| = 2\sqrt {M({\rho _{AB}})},
    	  \end{equation} 
    	 where  $M({\rho _{AB}}) = {\max _{i < k}}({u_i} + {u_k}) \le 2$ and ${u_i}(i = 1,2,3)$ are the eigenvalues of the symmetric matrix $U = {T^T}T$  constructed from the correlation matrix $T$  and its transpose ${T^T}$. The CHSH inequality can be violated if and only if (iff) $M({\rho _{AB}}) > 1$  \cite{w19}. In order to quantify the maximal violation of the CHSH inequality, we can use  $M({\rho _{AB}})$ or, equivalently \cite{w47}
           \begin{equation}\label{E06}  
                N({\rho _{AB}}) \equiv \max \left\{ {0,{\rm{ 2}}\sqrt {M({\rho _{AB}})}  - 2} \right\},
           \end{equation} 
         which yields $N({\rho _{AB}}) = 0$  if the CHSH inequality is not violated and $N({\rho _{AB}}) = 2\sqrt2-2$  for its maximal violation. In addition, with respect to a two-qubit \textit{X}-state, the three eigenvalues ${\mu _i}$  of the real symmetric matrix  $U = {T^T}T$ are \cite{w03,w14}
           \begin{equation}\label{E07}  
                \begin{split}
                &{\mu _1} = 4{\left( {\left| {{\rho _{14}}} \right| + \left| {{\rho _{23}}} \right|} \right)^2},\\
                &{\mu _2} = 4{\left( {\left| {{\rho _{14}}} \right| - \left| {{\rho _{23}}} \right|} \right)^2},\\
                &{\mu _3} = {\left( {{\rho _{11}} - {\rho _{22}} - {\rho _{33}} + {\rho _{44}}} \right)^2},
                \end{split}
           \end{equation}
         respectively, ${\rho _{ij}}(i,{\rm{ }}j = 1,{\rm{ }}2,{\rm{ }}3,{\rm{ }}4)$  are the matrix elements for two-qubit \textit{X}-state. It is easy to see that ${\mu _1}$  is always larger than ${\mu _2}$ , and thus the quantified CHSH inequality named Bell nonlocality can be expressed as
           \begin{equation}\label{E08}  
            N({\rho _{AB}}) = \max \left\{ {0,{\rm{ }}{B_1},{\rm{ }}{B_2}} \right\},
           \end{equation}
         here, ${B_1} = 2\sqrt {{\mu _1} + {\mu _2}}  - 2$ , and ${B_2} = 2\sqrt {{\mu _1} + {\mu _3}}  - 2$ .
     
     \section{The intrinsic relations among these QRs for different bipartite subsystems of the three-qubit systems} \label{sec3}
	    In this section, we mainly investigate the internal relations among these QRs in the case of a qubit-qubit subsystem for the multiparticle systems. Here, the multiparticle system is a three-qubit \textit{W}-type state. Labeling the qubits with 1, 2, and 3, the condition  $\left\langle {\hat n} \right\rangle  = \left\langle {{{\hat n}_1}} \right\rangle  + \left\langle {{{\hat n}_2}} \right\rangle  + \left\langle {{{\hat n}_3}} \right\rangle  = 1$ identifies the states whose wave function can be written in the form
	       \begin{equation}\label{E09}  
	       \left| \varphi  \right\rangle _{123}^{W - c} = \alpha \left| {001} \right\rangle  + \beta \left| {010} \right\rangle  + \gamma \left| {100} \right\rangle,
	      \end{equation}
	   where  ${\alpha ^2} + {\beta ^2} + {\gamma ^2} = 1$. The corresponding density matrix can be expressed as
	    \begin{equation}\label{E10}  
	    \begin{split}
	     \rho _{123}^{W - c} = &\gamma {\alpha ^*}\left| {100} \right\rangle \left\langle {001} \right| + \gamma {\beta ^*}\left| {100} \right\rangle \left\langle {010} \right|\\
	     &{\rm{          }} + \alpha {\beta ^*}\left| {001} \right\rangle \left\langle {010} \right| + \alpha {\gamma ^*}\left| {001} \right\rangle \left\langle {100} \right|\\
	     &{\rm{          }} + \beta {\alpha ^*}\left| {010} \right\rangle \left\langle {001} \right| + \beta {\gamma ^*}\left| {010} \right\rangle \left\langle {100} \right|{\rm{ }}\\
	     &{\rm{          }} + {\alpha ^2}\left| {001} \right\rangle \left\langle {001} \right| + {\beta ^2}\left| {010} \right\rangle \left\langle {010} \right|\\
	     &{\rm{          }} + {\gamma ^2}\left| {100} \right\rangle \left\langle {100} \right|.
	   \end{split}
	   \end{equation}
	   Considering different pairs of qubits described by the partially reduced density matrix from $\rho _{123}^{W - c}$ in Eq. (\ref{E10}), by utilizing Eqs. (\ref{E02}), (\ref{E03}) and (\ref{E08}), the corresponding DC, concurrence, and Bell nonlocality can be expressed as follows
	   \begin{equation}\label{E11}  
	   \begin{split}
	   &D_{13}^2 = 2\left( {{\alpha ^4} - {\alpha ^2} + {\gamma ^4} - {\gamma ^2}} \right) + 1,\\
	   &D_{12}^2 = 2\left( {{\beta ^4} - {\beta ^2} + {\gamma ^4} - {\gamma ^2}} \right) + 1,\\
	   &D_{23}^2 = 2\left( {{\alpha ^4} - {\alpha ^2} + {\beta ^4} - {\beta ^2}} \right) + 1,
	   \end{split}
	   \end{equation}	   
	   \begin{equation}\label{E12}  
	   \begin{split}
	   &{C_{13}} = 2\sqrt {{\alpha ^2}{\gamma ^2}} ,\\
	   &{C_{12}} = 2\sqrt {{\beta ^2}{\gamma ^2}} ,\\
	   &{C_{23}} = 2\sqrt {{\alpha ^2}{\beta ^2}} ,
	   \end{split}
	   \end{equation}
	   and
	   \begin{equation}\label{E13}  
	  \begin{split}
	    N_{13}^{{B_{\max }}} =&\max \left\{ {0,{\rm{ }}4\sqrt {2{\alpha ^2}{\gamma ^2}}  - 2,} \right.\\
	    &{\rm{ }}\left. {2\sqrt {4{\alpha ^2}{\gamma ^2} + {{(2{\alpha ^2} + 2{\gamma ^2} - 1)}^2}}  - 2} \right\},\\
	    N_{12}^{{B_{\max }}} =& \max \left\{ {0,{\rm{ }}4\sqrt {2{\beta ^2}{\gamma ^2}}  - 2,} \right.\\
	    &\left. {{\rm{ }}2\sqrt {4{\beta ^2}{\gamma ^2} + {{(2{\beta ^2} + 2{\gamma ^2} - 1)}^2}}  - 2} \right\},\\
	    N_{23}^{{B_{\max }}} =& \max \left\{ {0,{\rm{ }}4\sqrt {2{\alpha ^2}{\beta ^2}}  - 2,} \right.\\
	    &\left. {{\rm{ }}2\sqrt {4{\alpha ^2}{\beta ^2} + {{(2{\alpha ^2} + 2{\beta ^2} - 1)}^2}}  - 2} \right\},
	  \end{split}
	  \end{equation}
	   respectively. The conditions of determining the generation of Bell nonlocality can be deduced from Eq. (\ref{E13}), which do not yield Bell nonlocality if $N({\rho _{AB}}) = 0$, and $N({\rho _{AB}})=2\sqrt2-2$ for the maximal values of Bell nonlocality.
	   \begin{figure} 
	   	\centering
	   	\includegraphics[scale=0.9]{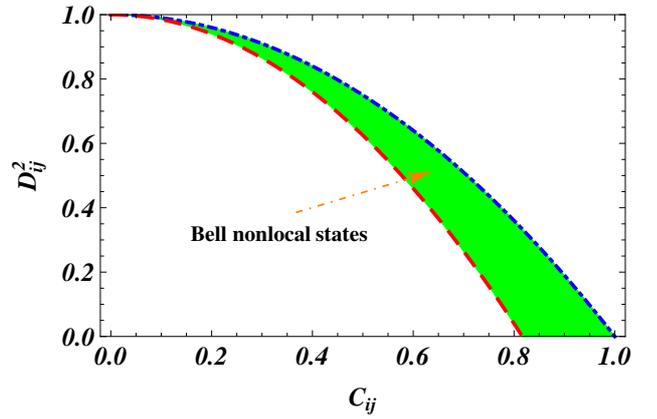}
	   	\caption{ \small The DC $D_{ij}^2$  versus the concurrence $C_{ij}$ for BNBSs derived from the tripartite \textit{W}-type states  $\rho _{123}^{W-c}$. BNBSs are found in green areas. Red and blue curves are plotted in the light of the corresponding boundary formulas.}
	   	\label{Fig.1}
	   \end{figure}
       \begin{figure*}
        		\begin{minipage}[t]{0.5\linewidth}
        			\centering
        			\includegraphics[scale=0.9]{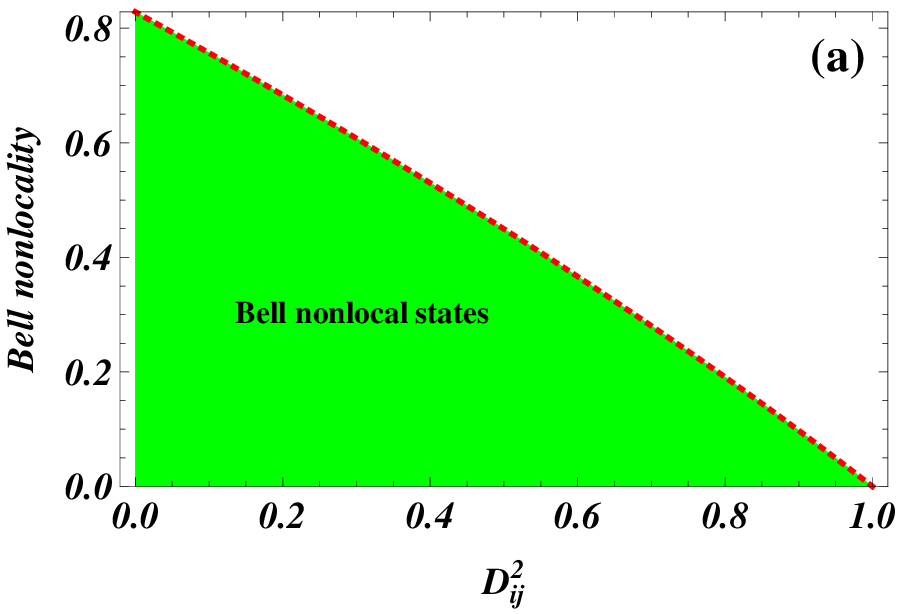}
        		\end{minipage}%
        		\begin{minipage}[t]{0.5\linewidth}
        			\centering
        			\includegraphics[scale=0.9]{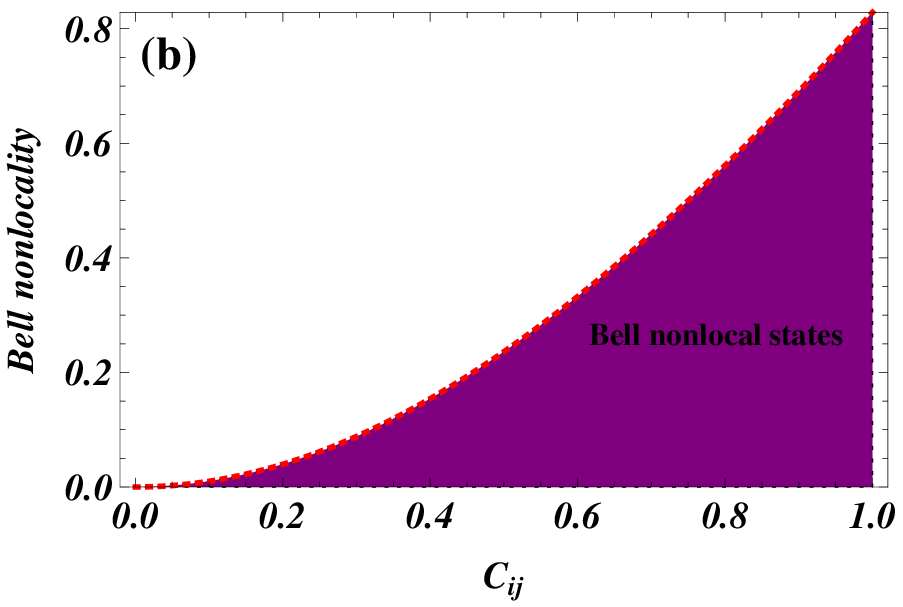}
        		\end{minipage}
        		\caption{ \small \textbf{(a)} The Bell nonlocality  $N_{ij}^{{B_{\max }}}$ versus the DC  $D_{ij}^2$ for BNBSs depending on Eq. (19). \textbf{(b)} The Bell nonlocality $N_{ij}^{{B_{\max }}}$  versus the concurrence ${C_{ij}}$   for BNBSs according to Eq. (\ref{E17}).}
        		\label{Fig.2}
        \end{figure*}
               
	   In our investigations, we firstly pay attention to the relation between the DC $D_{ij}^2$  and the concurrence ${C_{ij}}$ for Bell nonlocal bipartite states (BNBSs) described by the density matrix $\rho _{123}^{W - c}$  given in Eq. (\ref{E10}). The diagram depicting mutual relations between the DC and the concurrence is drawn in FIG. \ref{Fig.1}. The BNBSs are revealed in the green area. To discover the boundary conditions for the BNBSs plotted by blue dashed-dotted and red dashed curves, we should start with the expression of the DC $D_{ij}^2$  in Eq. (\ref{E11}), then, by applying Eqs. (\ref{E12}) and (\ref{E13}), we can obtain the boundary formed by BNBSs with the lowest possible DC assuming a fixed concurrence. Through numerical analysis, one can disclose that Bell nonlocality $N_{ij}^{{B_{\max }}}$  equals to zero for these bipartite states in the tripartite \textit{W}-type states. Via substitution of Eqs. (\ref{E11}) and (\ref{E12}) into Eq. (\ref{E13}), we can derive a formula as
	   \begin{equation}\label{E14}  
	   D_{ij}^2 + \frac{3}{2}C_{ij}^2 = 1.
	   \end{equation}
	   Besides, another boundary giving the maximal attainable DC for the fixed concurrence of BNBSs contains all bipartite pure states, where the DC can be written as follows
	   \begin{equation}\label{E15} 
	    \begin{split}
	    &D_{13}^2 = 2\left( {{\alpha ^4} + {\gamma ^4}} \right) - 1,\\
	    &D_{12}^2 = 2\left( {{\beta ^4} + {\gamma ^4}} \right) - 1,\\
	    &D_{23}^2 = 2\left( {{\alpha ^4} + {\beta ^4}} \right) - 1,
	    \end{split} 
	   \end{equation}
	   utilizing Eqs. (\ref{E11}) and (\ref{E15}), we can discover the upper boundary in FIG.  \ref{Fig.1} for the BNBSs obtaining following formula
	   \begin{equation}\label{E16} 
	    D_{ij}^2 + C_{ij}^2 = 1.
	   \end{equation}
	   By combining Eqs. (\ref{E14}) and (\ref{E16}), we can find that the DC for the BNBSs is located in an interval $D_{ij}^2 \in \left( {1 - \frac{3}{2}C_{ij}^2,{\rm{ }}1 - C_{ij}^2} \right)$ , which means that the DC and the concurrence have a trade-off relation: with the increasing of the concurrence, the DC will decrease, \textit{i.e.}, the concurrence enhances at the expense of DC. Note that this condition is similar to that proposed by Svozilik \textit{et al.} \cite{w35} for the parameter  ${S_{i,{\rm{ }}j}}$ (\textit{accessible coherence}) describing the maximal violation of the CHSH inequality $B_{ij}^{\max }$  and the DC, \textit{i.e.}, ${S_{i,{\rm{ }}j}} = {{D_{i,{\rm{ }}j}^2} \mathord{\left/{\vphantom {{D_{i,{\rm{ }}j}^2} 2}} \right.\kern-\nulldelimiterspace} 2} + {\left( {{{B_{ij}^{\max }} \mathord{\left/{\vphantom {{B_{ij}^{\max }} {2\sqrt 2 }}} \right.\kern-\nulldelimiterspace} {2\sqrt 2 }}} \right)^2}$. In addition, for the bipartite pure states, we can derive the relation between Bell nonlocality and the concurrence
	   \begin{equation}\label{E17} 
	    N_{ij}^{{B_{\max }}} = 2\sqrt {1 + C_{ij}^2}  - 2,
	   \end{equation}
	   due to $D_{ij}^2 + C_{ij}^2 = 1$, we can attain the relation between Bell nonlocality and the DC as
	   \begin{equation}\label{E18} 
	    D_{ij}^2 + 2{\left( {\frac{{N_{ij}^{{B_{\max }}} + 2}}{{2\sqrt 2 }}} \right)^2} = 2,
	   \end{equation}
	   namely,
	   \begin{equation}\label{E19} 
	    N_{ij}^{{B_{\max }}} = 2\sqrt {2 - D_{ij}^2} - 2,
	   \end{equation}
	   
	   Additionally, Bell nonlocality can be revealed only for bipartite entangled states with sufficiently strong nonlocal correlations. Thus, we can see that increasing the value of DC should be accompanied by decreasing the value of Bell nonlocality. Enhancing the value of the concurrence will synchronously increase Bell nonlocality. These results can really been disclosed for BNBSs in the randomly generated ensemble of states, as demonstrated in the green (FIG. \ref{Fig.2} (a)) and purple (FIG.\ref{Fig.2} (b)) areas, respectively. Furthermore, for a pure \textit{N}-party system, the entanglement of one party with the remaining $N-1$  parties confirms the purity of that party's quantum state when the rest of the quantum system is traced out. Actually, when a quantum system is a mixed state, it is so because of its entanglement with parties not sufficiently taken into account \cite{w48,w49,w50}. The purity of each party following the tracing out of the remaining ones can then regard as the basis for characterizing entanglement in multiparty systems. We know that the standard definition of purity for the bipartite states is
	   \begin{equation}\label{E20} 
	   {P_{AB}} = {\rm{Tr}}(\rho _{AB}^2).
	   \end{equation}
	   Thus, one can obtain the purity of the above bipartite states in the  tripartite \textit{W}-type states
	   \begin{equation}\label{E21} 
	   \begin{split}
	   &{P_{13}} = {\alpha ^4} + {\beta ^4} + 2{\alpha ^2}{\gamma ^2} + {\gamma ^4},\\
	   &{P_{12}} = {\alpha ^4} + {\beta ^4} + 2{\beta ^2}{\gamma ^2} + {\gamma ^4},\\
	   &{P_{23}} = {\alpha ^4} + {\beta ^4} + 2{\alpha ^2}{\beta ^2} + {\gamma ^4}.
	   \end{split} 
	   \end{equation}
	   Because of  ${\alpha ^2} + {\beta ^2} + {\gamma ^2} = 1$, by means of combining Eqs. (\ref{E11}), (\ref{E12}) and (\ref{E21}), an interesting result can be derived
	   \begin{equation}\label{E22} 
	    D_{ij}^2 + C_{ij}^2 = {P_{ij}}.
	   \end{equation}
	   Because the purity is equal to one iff the bipartite state is a pure state, we can draw a conclusion
	   \begin{equation}\label{E23} 
	    D_{ij}^2 + C_{ij}^2 \le 1.
	   \end{equation}
	   The inequality (\ref{E23}) also illustrates that the concurrence increases at the expense of the degree of coherence in the bipartite subsystems derived the  tripartite \textit{W}-type states.
	   
	  \begin{figure}
	   \centering
	   \includegraphics[scale=0.9]{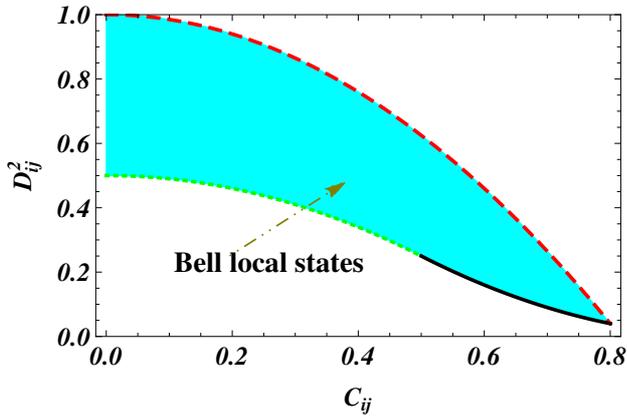}
	   \caption{ \small The degree of coherence $D_{ij}^2$  versus the concurrence ${C_{ij}}$  for BLBSs derived from the  tripartite \textit{W}-type states  $\rho _{123}^{W-c}$. BLBSs are discovered in cyan areas. Red dashed, green dashed and black solid curves are plotted in accordance with the corresponding boundary formulas, respectively.}
	  \label{Fig.3}
      \end{figure}

	   Subsequently, we are engaged in the achieved boundaries of Bell local bipartite states (BLBSs) drawn in FIG. \ref{Fig.3}. For ${C_{ij}} > {4 \mathord{\left/{\vphantom {4 5}} \right.\kern-\nulldelimiterspace}5}$ , the maximal attainable DC of BLBSs is given by Eq. (\ref{E14}), which is already obtained for BNBSs. Besides, coinstantaneous discussion for the DC of Eq. (\ref{E11}) and the concurrence of Eq. (\ref{E12}), affords us the minimal accessible DC in the scope of  ${{{C_{ij}} \ge 1} \mathord{\left/{\vphantom {{{C_{ij}} \ge 1} 2}} \right.\kern-\nulldelimiterspace} 2}$:
	   \begin{equation}\label{E24} 
	    D_{ij}^2 - C_{ij}^2 + 2{C_{ij}} = 1.
	   \end{equation}
	   \begin{center}
	   	\begin{minipage}{245pt}
	   		{\linespread{0.2}{\small \textbf{Table 1}. Revealed BNBSs and BLBSs in the area $\left( {D_{ij}^2,{\rm{ }}C_{ij}^2} \right)$  stepped over the degree of coherence and the concurrence.}}
	   		\begin{center}
	   			\begin{tabular}{|l|l|l|}
	   				\hline \small Concurrence & \small The degree of coherence & \small Revealed states\\  \hline
	   				\small	$0 < {C_{ij}} \le \frac{1}{2}$ & \small$1 - \frac{3}{2}C_{ij}^2 < D_{ij}^2 \le 1 - C_{ij}^2$ & \small Only BNBSs \\
	   				\small	$0 < {C_{ij}} \le \frac{1}{2}$  &\small $\frac{1}{2} - C_{ij}^2 \le D_{ij}^2 \le 1 - \frac{3}{2}C_{ij}^2$ & \small Only BLBSs \\
	   				\small	$\frac{1}{2} < {C_{ij}} \le \frac{4}{5}$  & \small$1 - \frac{3}{2}C_{ij}^2 < D_{ij}^2 \le 1 - C_{ij}^2$ & \small Only BNBSs \\
	   				\small	$\frac{1}{2} < {C_{ij}} \le \frac{4}{5}$ & \small${\left( {1 - {C_{ij}}} \right)^2} \le D_{ij}^2 \le 1 - \frac{3}{2}C_{ij}^2$ & \small Only BLBSs\\
	   				\small	$\frac{4}{5} < {C_{ij}} \le 1$ &\small $1 - \frac{3}{2}C_{ij}^2 < D_{ij}^2 \le 1 - C_{ij}^2$ & \small Only BNBSs  \\
	   				\hline
	   			\end{tabular}\\
	   		\end{center}\label{tab1}	
	   	\end{minipage}
	   \end{center}\vskip.2cm
	   Here, the boundary formula (\ref{E24}) corresponds to the black solid curve in FIG. \ref{Fig.3}, and the underlying states are the Horodecki states that are composed of a Bell state and a vacuum state. The Horodecki state can be written as
	   \begin{equation}\label{E25} 
	    {\rho _H} = \varepsilon \left| {{\phi ^ + }} \right\rangle \left\langle {{\phi ^ + }} \right| + (1 - \varepsilon )\left| {00} \right\rangle \left\langle {00} \right|,
	   \end{equation}
	   where $\varepsilon$  is the state parameter (run from 0 to 1) which gives different prepared states, and $\left| {{\phi ^ + }} \right\rangle  = \frac{1}{{\sqrt 2 }}\left( {\left| {01} \right\rangle  + \left| {10} \right\rangle } \right)$. Here we can attain the concurrence, the DC, purity and Bell nonlocality
	  \begin{equation}\label{E26} 
	   \begin{split}
	   {C_{({\rho _H})}}= &\varepsilon ,{\rm{ }}D_{({\rho _H})}^2 = {\left( {1 - \varepsilon } \right)^2},\\
	   {P_{({\rho _H})}} = &{\left( {1 - \varepsilon } \right)^2} + {\varepsilon ^2},\\
	   N_{({\rho _H})}^{{B_{\max }}} = &\max \left\{ {0,{\rm{ }}2\sqrt {2{\varepsilon ^2}}  - 2,} \right.\\
	   &{\rm{            }}\left. {2\sqrt {1 + \varepsilon \left( {5\varepsilon  - 4} \right)}  - 2} \right\},
	   \end{split}   
	  \end{equation}
	   respectively. Hence, we can draw the same conclusion: the DC plus the square of the concurrence is always equal to the purity: $D_{ij}^2 + C_{ij}^2 = {P_{ij}}$. Note that, EPR steering can be disclosed in the Horodecki states if we use the linear criterion of EPR steering introduced by Cavalcanti, Jones, Wiseman, and Reid (CJWR) \cite{w51} in 2009. Consequently, the some results of Ref. \cite{w38} are incomplete and less precise, one cannot say that the state formulated by Eq. (\ref{E25}) is an unsteerable state even though the state cannot violate the specific EPR steering inequality \cite{w52}. The last boundary corresponding to the green dashed curve in FIG. \ref{Fig.3} giving the minimal degree of coherence for the corresponding the concurrence ${{{C_{ij}} < 1} \mathord{\left/{\vphantom {{{C_{ij}} < 1} 2}} \right. \kern-\nulldelimiterspace}2}$  is given by
	   \begin{equation}\label{E27} 
	    D_{ij}^2 + C_{ij}^2 = \frac{1}{2},
	   \end{equation}
	   and the underlying states have the following density matrix
	   \begin{equation}\label{E28} 
	   \begin{split}
		   {\rho _{ij}} = &\frac{1}{2}\left| {00} \right\rangle \left\langle {00} \right| + \left( {{1 \mathord{\left/
	   			{\vphantom {1 2}} \right. \kern-\nulldelimiterspace} 2} - a} \right)\left| {01} \right\rangle \left\langle {01} \right|\\
	       &{\rm{  }} + \sqrt {{a \mathord{\left/{\vphantom {a 2}} \right.
	   			\kern-\nulldelimiterspace} 2} - {a^2}} \left( {\left| {01} \right\rangle \left\langle {10} \right| + \left| {10} \right\rangle \left\langle {01} \right|} \right)\\
	       &{\rm{ }} + a\left| {10} \right\rangle \left\langle {10} \right|.
	    \end{split} 
	   \end{equation}
	   The expression of DC, concurrence, purity and Bell nonlocality can be obtained, and $D_{ij}^2 + C_{ij}^2 = {P_{ij}} \le 1$ can be also unveiled. The discussion of the attainable DC and concurrence is showed in the diagrams of FIGs. \ref{Fig.1} and \ref{Fig.3}, then, we can identify different areas in the plane $\left( {D_{ij}^2,{\rm{ }}C_{ij}^2} \right)$ from the perspective of Bell nonlocality or not. The analysis results are shown in Table 1. In accordance with the results, BNBSs and BLBSs can be revealed in the certain areas.
			   
    \section{Mutual relations between QRs and purity in two specific scenarios for a tripartite \textit{W}-type state} \label{sec4}
	    	In this section, we will consider two specific scenarios for a tripartite \textit{W}-type state. One case is a tripartite  \textit{W}-type state under decoherence channel, we mainly consider how the phase flip (PF) channel affects the mutual relations among these QRs.  And the other one is a renormalized spin-1/2 Heisenberg \textit{XXZ} model, some exact internal relations between QRs and purity are revealed in this practical system.
	    		    		  	   		
        \subsection{The tripartite \textit{W}-type state under PF channel}  \label{sec4a}
    		We will discuss the mutual relations of the DC, the concurrence, purity and Bell nonlocality for final states $\rho _{123}^{PF}$  (the tripartite \textit{W}-type states suffered from PF channel) described by a trace-preserving quantum operation  $\vartheta (\rho )$, which is given by $\vartheta (\rho ) = \sum\limits_{i = 1,2} {\left( {E_i^1 \otimes E_i^2 \otimes E_i^3} \right)}  \cdot \rho {\left( {E_i^1 \otimes E_i^2 \otimes E_i^3} \right)^\dag }$  where $\left\{ {{E_i}} \right\}$  is the set of Kraus operators associated to a decohering process of a single qubit, with the trace-preserving condition reading $\sum\nolimits_i {E_i^\dag {E_i}}  = I$ \cite{w53}. The Kraus operators of PF channel can be expressed as \cite{w14,w54}
    	\begin{equation}\label{E29} 
    		{E_1} = \sqrt p  {I_2},{\rm{ }}{E_2} = \sqrt {1 - p} {\sigma _z},
    	\end{equation}
    		where  ${I_2}$ is a $2\times2$ identity matrix, ${\sigma _z}$  is a Pauli matrix at site \textit{z}, and $0 \le p = 1 - {e^{ - \eta t}} \le 1$  is the PF channel decoherence strength and $\eta $  is the decay rate. Then, we can obtain the final states
    	\begin{equation}\label{E30} 
    		\begin{split}
    		\rho _{123}^{PF} = &\gamma Y{\alpha ^*}\left| {100} \right\rangle \left\langle {001} \right| + \gamma Y{\beta ^*}\left| {100} \right\rangle \left\langle {010} \right|\\
    		&{\rm{        }} + \alpha Y{\beta ^*}\left| {001} \right\rangle \left\langle {010} \right| + \alpha Y{\gamma ^*}\left| {001} \right\rangle \left\langle {100} \right|\\
    		&{\rm{        }} + \beta Y{\alpha ^*}\left| {010} \right\rangle \left\langle {001} \right| + \beta Y{\gamma ^*}\left| {010} \right\rangle \left\langle {100} \right|{\rm{ }}\\
    		&{\rm{        }} + {\alpha ^2}\left| {001} \right\rangle \left\langle {001} \right| + {\beta ^2}\left| {010} \right\rangle \left\langle {010} \right|\\
    		&{\rm{        }} + {\gamma ^2}\left| {100} \right\rangle \left\langle {100} \right|,
    	    \end{split} 
    	\end{equation}
    		here, $Y = {\left( {1 - 2p} \right)^2}$. By separately tracing over the qubit 3, 2 and 1, we can obtain that the reduced density matrices are 
    	\begin{subequations}   
    	  \begin{align}
    	    &\rho _{12}^{PF} = \left( {\begin{array}{*{20}{c}}
    	    	{{\alpha ^2}}&0&0&0\\
    	    	0&{{\beta ^2}}&{Y\beta {\gamma ^*}}&0\\
    	    	0&{Y{\beta ^*}\gamma }&{{\gamma ^2}}&0\\
    	    	0&0&0&0
    	    	\end{array}} \right),\label{E31a}\\
        	&\rho _{13}^{PF} = \left( {\begin{array}{*{20}{c}}
    	    	{{\beta ^2}}&0&0&0\\
    	    	0&{{\alpha ^2}}&{Y\alpha {\gamma ^*}}&0\\
    	    	0&{Y{\alpha ^*}\gamma }&{{\gamma ^2}}&0\\
    	    	0&0&0&0
    	    	\end{array}} \right),\label{E31b}\\
            &\rho _{23}^{PF} = \left( {\begin{array}{*{20}{c}}
    	    	{{\gamma ^2}}&0&0&0\\
    	    	0&{{\alpha ^2}}&{Y\alpha {\beta ^*}}&0\\
    	    	0&{Y{\alpha ^*}\beta }&{{\beta ^2}}&0\\
    	    	0&0&0&0
    	    	\end{array}} \right),\label{E31c}
    	   \end{align}
    	\end{subequations}
    	respectively. Via employing Eqs. (\ref{E02}), (\ref{E03}), (\ref{E08}) and (\ref{E20}), the corresponding purity, concurrence, DC and Bell nonlocality can be expressed as follows
    	\begin{equation}\label{E32}  
    		\begin{split}
    		&{P_{13}}(PF) = {\alpha ^4} + {\beta ^4} + 2{Y^2}{\alpha ^2}{\gamma ^2} + {\gamma ^4},\\
    		&{P_{12}}(PF) = {\alpha ^4} + {\beta ^4} + 2{Y^2}{\beta ^2}{\gamma ^2} + {\gamma ^4},\\
    		&{P_{23}}(PF) = {\alpha ^4} + {\beta ^4} + 2{Y^2}{\alpha ^2}{\beta ^2} + {\gamma ^4},
    		\end{split}
    	\end{equation}
    	\begin{equation}\label{E33}  
    		\begin{split}
    		&{C_{13}}(PF) = 2\sqrt {{Y^2}{\alpha ^2}{\gamma ^2}} ,{\rm{ }}\\
    		&{C_{12}}(PF) = 2\sqrt {{Y^2}{\beta ^2}{\gamma ^2}} ,{\rm{ }}\\
    		&{C_{23}}(PF) = 2\sqrt {{Y^2}{\alpha ^2}{\beta ^2}} ,
    		\end{split}
    	\end{equation}
    	\begin{equation}\label{E34}  
    	  \begin{split}
    	  &D_{13}^2(PF) = 2\left( {{\alpha ^4} - {\alpha ^2} + {\gamma ^4} - {\gamma ^2}} \right) + 1,\\
    	  &D_{12}^2(PF) = 2\left( {{\beta ^4} - {\beta ^2} + {\gamma ^4} - {\gamma ^2}} \right) + 1,\\
    	  &D_{23}^2(PF) = 2\left( {{\alpha ^4} - {\alpha ^2} + {\beta ^4} - {\beta ^2}} \right) + 1,
    	  \end{split}
    	\end{equation}
    	and
    	\begin{equation}\label{E35}  
    	  \begin{split}
    	  N_{13}^{{B_{\max }}}&(PF) = \max \left\{ {0,{\rm{ }}4\sqrt {2{Y^2}{\alpha ^2}{\gamma ^2}}  - 2,{\rm{ }}} \right.\\
    	  &\left. {{\rm{       }}2\sqrt {4{Y^2}{\alpha ^2}{\gamma ^2} + {{(2{\alpha ^2} + 2{\gamma ^2} - 1)}^2}}  - 2} \right\},\\
    	  N_{12}^{{B_{\max }}}&(PF)  =\max \left\{ {0,{\rm{ }}4\sqrt {2{Y^2}{\beta ^2}{\gamma ^2}}  - 2,{\rm{ }}} \right.\\
    	  &{\rm{      }}\left. {2\sqrt {4{Y^2}{\beta ^2}{\gamma ^2} + {{(2{\beta ^2} + 2{\gamma ^2} - 1)}^2}}  - 2} \right\},\\
    	  N_{23}^{{B_{\max }}}&(PF) = \max \left\{ {0,{\rm{ }}4\sqrt {2{Y^2}{\alpha ^2}{\beta ^2}}  - 2,} \right.\\
    	  &\left. {{\rm{      }}2\sqrt {4{Y^2}{\alpha ^2}{\beta ^2} + {{(2{\alpha ^2} + 2{\beta ^2} - 1)}^2}}  - 2} \right\},
    	 \end{split}
    	\end{equation}	
    	respectively.		
    	
    	\begin{figure*}
    		\begin{minipage}[t]{0.5\linewidth}
    			\centering
    			\includegraphics[scale=0.9]{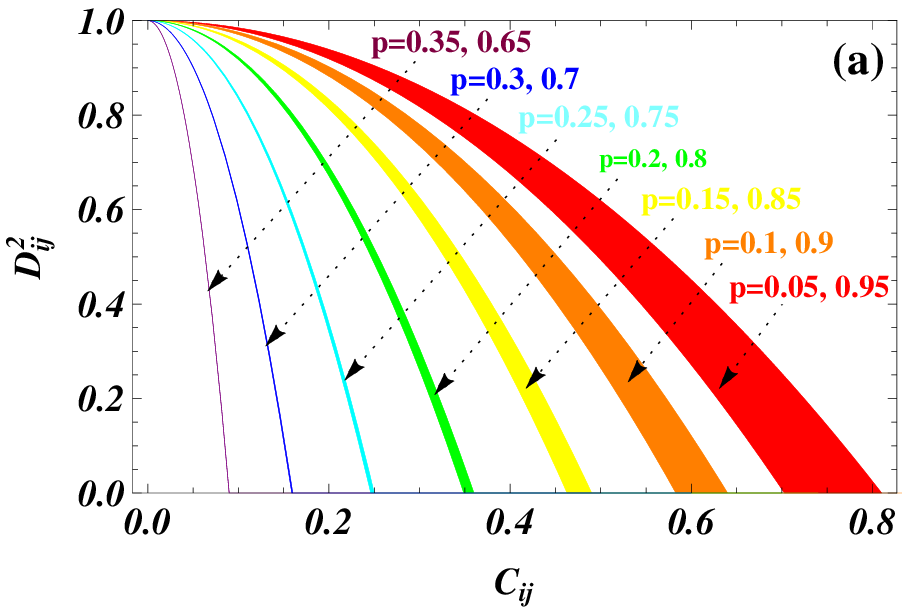}
    		\end{minipage}%
    		\begin{minipage}[t]{0.5\linewidth}
    			\centering
    			\includegraphics[scale=0.9]{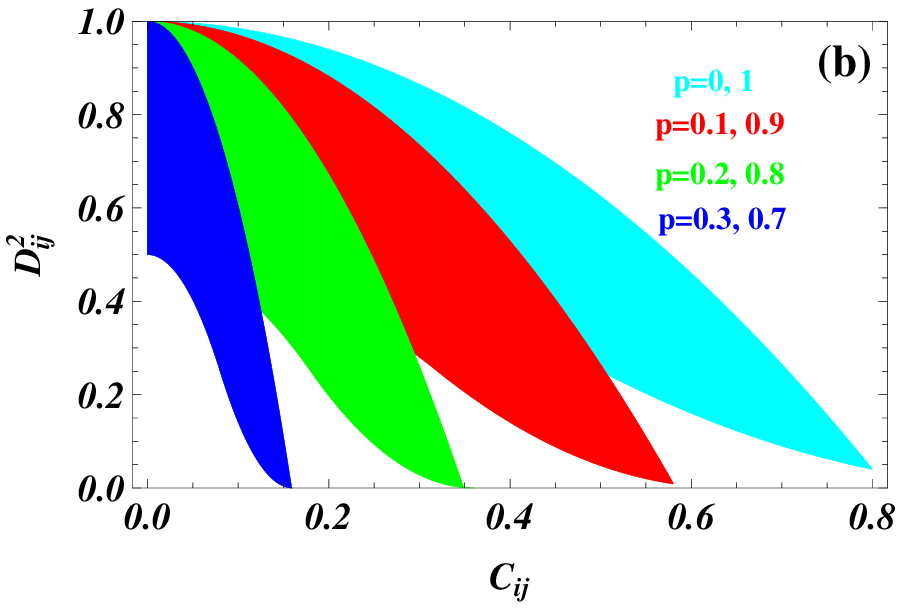}
    		\end{minipage}
    		\caption{ \small  The degree of coherence versus the concurrence for Bell nonlocal \textbf{(a)} and Bell local \textbf{(b)} bipartite states derived from the tripartite \textit{W}-type states in the PF channel $\rho _{123}^{PF}$ given in Eq. (\ref{E30}). Bell nonlocal (Bell local) states are discovered in different rainbow colors areas (cyan, red, green and blue areas) corresponding to different decoherence strength \textit{p}. }
    		\label{Fig.4} 
    	\end{figure*}
    
    	Now, we focus on the relation between the DC and the concurrence for BNBSs. The diagram depicting mutual relations between the DC and the concurrence is plotted in FIG. \ref{Fig.4}. The BNBSs can be discovered in the different rainbow colors areas corresponding to the different decoherence strength in FIG. \ref{Fig.4} (a), whereas BLBSs are disclosed in the cyan, red, green and blue areas corresponding to the different decoherence strength in FIG. \ref{Fig.4} (b). To find the lower boundary conditions for the BNBSs in FIG. \ref{Fig.4} (a), we use the expression of Bell nonlocality in Eq. (\ref{E35}), then, from Eqs. (\ref{E33}) and (\ref{E34}), we analyze the lower boundary formed by BNBSs with the lowest possible the DC assuming a fixed concurrence. Through analysis, Bell nonlocality should be equal to zero for these bipartite states in the tripartite states $\rho _{123}^{PF}$. Then, we can obtain a wishful lower boundary formula
    	\begin{equation}\label{E36}  
          D_{ij}^2 + \frac{{C_{ij}^2\left( {2 + {Y^2}} \right)}}{{2{Y^2}}} = 1.
    	\end{equation}
    	The upper boundary condition for the BNBSs in FIG. \ref{Fig.4} (a) giving the maximal attainable DC for the fixed concurrence of BNBSs for the bipartite pure states undergone the PF channel, which the corresponding DC can be obtained. By utilizing prior method, we can reveal the upper boundary condition for the BNBSs as follow
    	\begin{equation}\label{E37}  
    	 D_{ij}^2 + \frac{{C_{ij}^2}}{{{Y^2}}} = 1.
    	\end{equation}
    	
    	Next, we start with the achieved boundaries of BLBSs in FIG. \ref{Fig.4} (b). The maximal attainable DC of BLBSs is given by the upper boundary Eq. (\ref{E36}) already obtained for BNBSs. In addition, coinstantaneous discussion of the concurrence Eq. (\ref{E33}) and the DC Eq. (\ref{E34}), respectively, affords us the minimal accessible DC in the scope of  ${{{C_{ij}} \ge Y} \mathord{\left/{\vphantom {{{C_{ij}} \ge Y} 2}} \right.\kern-\nulldelimiterspace} 2}$, we can disclose a lower boundary for BLBSs
    	 \begin{equation}\label{E38}  
    		D_{ij}^2 = {\left( {1 - \frac{{{C_{ij}}}}{Y}} \right)^2}.
    	 \end{equation}
    	Here, the boundary condition (\ref{E38}) corresponds to the lower boundary for ${{{C_{ij}} \ge Y} \mathord{\left/{\vphantom {{{C_{ij}} \ge Y} 2}} \right.\kern-\nulldelimiterspace} 2}$  in FIG. \ref{Fig.4} (b), and the underlying states have all qubits for the Horodecki states formulated (\ref{E25}) undergone the PF channel. Then, another lower boundary giving the minimal DC for the corresponding concurrence ${{{C_{ij}} < Y} \mathord{\left/{\vphantom {{{C_{ij}} < Y} 2}} \right.\kern-\nulldelimiterspace} 2}$  is written as	
    	\begin{equation}\label{E39}  
    		D_{ij}^2 + \frac{{C_{ij}^2}}{{{Y^2}}} = \frac{1}{2},
    	\end{equation}
    	and the underlying states have the following density matrix	
    	\begin{equation}\label{E40}  
    	\begin{split}
    	\rho _{ij}^{PF} = &\frac{1}{2}\left| {00} \right\rangle \left\langle {00} \right| + \left( {{1 \mathord{\left/
    				{\vphantom {1 2}} \right.
    				\kern-\nulldelimiterspace} 2} - a} \right)\left| {01} \right\rangle \left\langle {01} \right|\\
    	&{\rm{         }} + Y\sqrt {{a \mathord{\left/
    				{\vphantom {a 2}} \right.
    				\kern-\nulldelimiterspace} 2} - {a^2}} \left( {\left| {01} \right\rangle \left\langle {10} \right| + \left| {10} \right\rangle \left\langle {01} \right|} \right)\\
    	&{\rm{         }} + a\left| {10} \right\rangle \left\langle {10} \right|.
    	\end{split}
    	\end{equation}	
    	As shown in FIG. \ref{Fig.4}, we can see that the proportion of the areas of BNBSs relative to BLBSs will decrease with the increase of decoherence strength. Besides, the concurrence decreases with the growing intensity of decoherence, whereas, the DC is not influenced by the decoherence, \textit{i.e.}, the DC is immune to the PF noise. Additionally, due to  ${\alpha ^2} + {\beta ^2} + {\gamma ^2} = 1$, we still get hold of	
    	 \begin{equation}\label{E41}  
    		D_{ij}^2 + C_{ij}^2 = {P_{ij}} \le 1,
    	 \end{equation}
    	\textit{i.e.}, the degree of coherence plus the square of the concurrence is always equal to the purity. And the concurrence enhances at the expense of the DC, which is a significant conclusion.
    		
     \subsection{A practical physical system for spin-1/2 Heisenberg \textit{XXZ} model}  \label{sec4b}
    	Now, we simply introduce a renormalized spin-1/2 Heisenberg \textit{XXZ} model. The Hamiltonian of spin-1/2 Heisenberg \textit{XXZ} model on a periodic chain of \textit{N} sites is \cite{w55, w56, w57}	
    	 \begin{equation}\label{E42} 
         H = \frac{J}{4}\sum\limits_{k = 1}^N {(\sigma _k^x\sigma _{k + 1}^x + \sigma _k^y\sigma _{k + 1}^y + \delta \sigma _k^z\sigma _{k + 1}^z)},
         \end{equation}	
    	where $\delta$ is the anisotropy parameter, $J$ is the exchange constant, and $J,{\rm{ }}\delta  > 0$. $\sigma _k^\beta (\beta  = x,y,z)$  is standard Pauli matrices at site \textit{k}. By employing the Kadanoff's block method, it is necessary to divide the initial system Hamiltonian shown in Eq. (\ref{E42}) into two parts \cite{w56, w57}
    	 \begin{equation}\label{E43} 
    	  H = {H^D} + {H^{DD}},
    	 \end{equation}	
    	where  ${H^D}$ is the block Hamiltonian and  ${H^{DD}}$ is the interblock Hamiltonian. The perturbative implementa-tion of this method has been discussed in Refs. \cite{w55, w58}. One can present this approach in the first-order correction. The effective Hamiltonian is given by \cite{w59}	
    		\begin{equation}\label{E44} 
    		{H^{eff}} = {P_0}{H^D}{P_0} + {P_0}{H^{DD}}{P_0},
    		\end{equation}
    	\begin{figure*}
    		\begin{minipage}[t]{0.5\linewidth}
    			\centering
    			\includegraphics[scale=0.9]{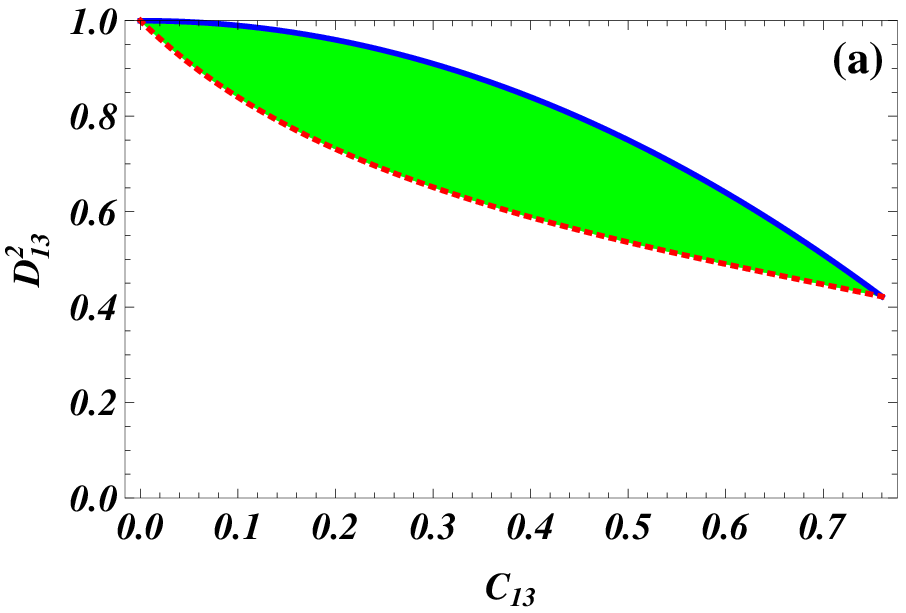}
    		\end{minipage}%
    		\begin{minipage}[t]{0.5\linewidth}
    			\centering
    			\includegraphics[scale=0.9]{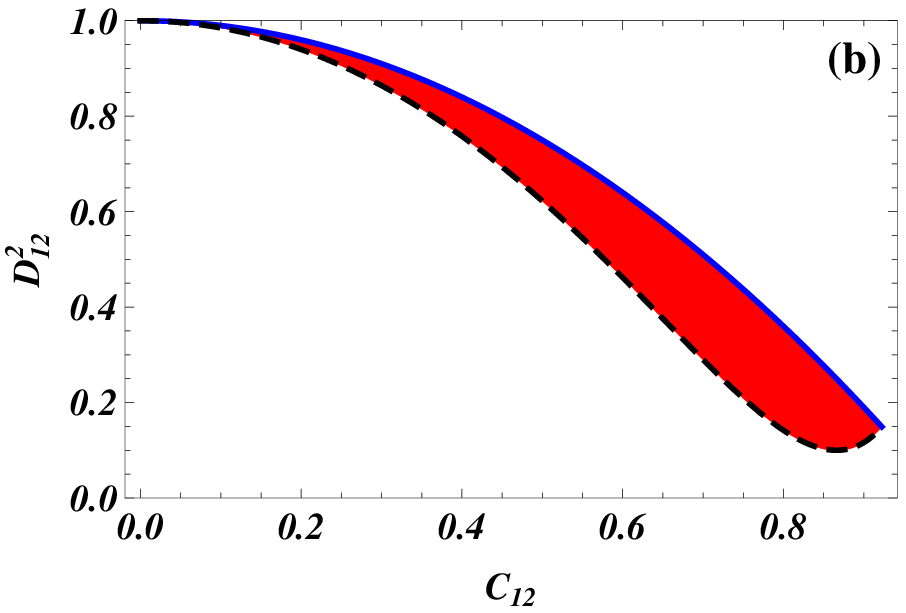}
    		\end{minipage}
    		\caption{  \small The degree of coherence versus the concurrence for BNBSs derived from the tripartite \textit{W}-type states in spin-1/2 \textit{XXZ} model given in Eq. (\ref{E47}). \textbf{(a)} BNBSs are discovered in green areas for the state  ${\rho _{13}}(XXZ)$, and \textbf{(b)} BNBSs are disclosed in red areas for the states ${\rho _{12}}(XXZ),{\rho _{23}}(XXZ)$. Blue solid, red dashed and black dashed curves are plotted in accordance with the corresponding boundary formulas, respectively.}
    		\label{Fig.5}
    	\end{figure*}	
    	where ${P_0}$  is a projection operator. To get a renormalized form of the Hamiltonian, one can use a three-site block procedure. Note that, choosing the three-site block is essential here to get a self-similar Hamiltonian after each quantum renormalization group step. Subsequently, according to Eq. (\ref{E44}), the effective Hamiltonian of the renormalized chain can be given by \cite{w56, w59}	
    		\begin{equation}\label{E45} 
    		{H^{{\rm{e}}ff}} = \frac{{J'}}{4}\sum\limits_{k = 1}^{N/3} {(\sigma _k^x\sigma _{k + 1}^x + \sigma _k^y\sigma _{k + 1}^y + \delta '\sigma _k^z\sigma _{k + 1}^z)}.
    		\end{equation} 
    	Hence, the iterative relations are
          \begin{equation}\label{E46} 
           J' = J{\left( {{{2q} \mathord{\left/
           				{\vphantom {{2q} {2 + {q^2}}}} \right.
           				\kern-\nulldelimiterspace} {2 + {q^2}}}} \right)^2},{\rm{ }}\delta ' = \delta {{{q^{\rm{2}}}} \mathord{\left/
           		{\vphantom {{{q^{\rm{2}}}} {\rm{4}}}} \right.
           		\kern-\nulldelimiterspace} {\rm{4}}}.
          \end{equation}
        Herein, the degenerate ground states are given by \cite{w59, w60}
            \begin{equation}\label{E47} 
             \left| {{\psi _0}} \right\rangle  = {({\rm{2}} + {q^{\rm{2}}})^{{{ - 1} \mathord{\left/
             				{\vphantom {{ - 1} 2}} \right.
             				\kern-\nulldelimiterspace} 2}}}\left\{ {\left| { \uparrow  \uparrow  \downarrow } \right\rangle  + q\left| { \uparrow  \downarrow  \uparrow } \right\rangle  + \left| { \downarrow  \uparrow  \uparrow } \right\rangle } \right\},    
            \end{equation}
    		\begin{equation}\label{E48} 
    		\left| {{\psi _0}'} \right\rangle  = {({\rm{2}} + {q^{\rm{2}}})^{ - {1 \mathord{\left/
    						{\vphantom {1 2}} \right.
    						\kern-\nulldelimiterspace} 2}}}\left\{ {\left| { \uparrow  \downarrow  \downarrow } \right\rangle  + q\left| { \downarrow  \uparrow  \downarrow } \right\rangle  + \left| { \downarrow  \downarrow  \uparrow } \right\rangle } \right\},    
    		\end{equation}
    	where $\left|  \uparrow  \right\rangle $  and $\left|  \downarrow  \right\rangle $  are the eigenstates of ${\sigma _z}$, and  $q =  - {{(\delta  + \sqrt {8 + {\delta ^2}} )} \mathord{\left/{\vphantom {{(\delta  + \sqrt {8 + {\delta ^2}} )} 2}} \right.\kern-\nulldelimiterspace} 2}$. The corresponding energy is given by ${E_0} =  - J{{(\delta  + \sqrt {8 + {\delta ^2}} )} \mathord{\left/{\vphantom {{(\delta  + \sqrt {8 + {\delta ^2}} )} 4}} \right.\kern-\nulldelimiterspace} 4}$. Then, the density matrix of the ground state can be read as  ${\rho _{123}}(XXZ) = \left| {{\psi _0}} \right\rangle \left\langle {{\psi _0}} \right|$, where  $\left| {{\psi _0}} \right\rangle $ is described as Eq. (\ref{E47}). Certainly, if we use  $\left| {{\psi _0}'} \right\rangle$ to construct the density matrix, trough calculations, the results will be the same ones. Therefore, we trace over site 1, 2 and 3 respectively, and obtain three different reduced density matrixes. In the same way, via using Eqs. (\ref{E02}), (\ref{E03}), (\ref{E08}) and (\ref{E20}), one can give the corresponding concurrence, DC, Bell nonlocality and purity as follows	
    	\begin{equation}\label{E49} 
    		\begin{split}
    		&{C_{13}}(XXZ) = \frac{2}{{2 + {q^2}}},{\rm{ }}\\
    		&{C_{12}}(XXZ) = {C_{23}}(XXZ) = \frac{{2\sqrt {{q^2}} }}{{2 + {q^2}}},{\rm{ }}
    		\end{split}   
    	\end{equation}
    	\begin{equation}\label{E50} 
    	   \begin{split}    
           &D_{13}^2(XXZ) = \frac{{{q^4}}}{{{{\left( {2 + {q^2}} \right)}^2}}},{\rm{ }}\\
           &D_{12}^2(XXZ) = D_{23}^2(XXZ) = \frac{{2 - 2{q^2} + {q^4}}}{{{{\left( {2 + {q^2}} \right)}^2}}},
       	   \end{split}   
    	\end{equation}	
    	\begin{equation}\label{E51} 
    	 \begin{split}    
            N_{13}^{{B_{\max }}}& (XXZ) = \max \left\{ {0,{\rm{ }}\frac{{2\sqrt {{q^4} - 4{q^2} + 8} }}{{2 + {q^2}}} - 2,} \right.\\
            &{\rm{    }}\left. {\frac{{4\sqrt 2 }}{{2 + {q^2}}} - 2} \right\},\\
            N_{12}^{{B_{\max }}}&(XXZ) =N_{23}^{{B_{\max }}}(XXZ)=\\
            &{\rm{   }} \max \left\{ {0,{\rm{ }}\frac{{4\sqrt {2{q^2}} }}{{2 + {q^2}}} - 2,} \right.
            {\rm{   }}\left. {\frac{{2\sqrt {{q^4} + 4{q^2}} }}{{2 + {q^2}}} - 2} \right\},
       	 \end{split}   
    	\end{equation}		
    	and
    	\begin{equation}\label{E52} 
    	\begin{split}    
        	&{P_{13}}(XXZ) = \frac{{4 + {q^4}}}{{{{\left( {2 + {q^2}} \right)}^2}}},{\rm{ }}\\
    	    &{P_{12}}(XXZ) = {P_{23}}(XXZ) = \frac{{2 + 2{q^2} + {q^4}}}{{{{\left( {2 + {q^2}} \right)}^2}}},
        \end{split}   
    	\end{equation}			
    	respectively.	
    	
    	The diagram depicting interrelations between the DC and the concurrence is plotted in FIG. \ref{Fig.5}. The BNBSs can be given in the green and red areas for the different bipartite states in the renormalized spin-1/2 Heisenberg \textit{XXZ} model in FIG. \ref{Fig.5}. The lower boundary conditions for the BNBSs of the state ${\rho _{13}}(XXZ)$  in FIG. \ref{Fig.5} (a), through analysis, Bell nonlocality should be equal to zero for the bipartite state, one can obtain the lower boundary formula as	
    	  \begin{equation}\label{E53} 
    	    D_{13}^2 = {\left( {\frac{{C_{13}^2 + \sqrt {C_{13}^2 - C_{13}^4} }}{{2C_{13}^2 + \sqrt {C_{13}^2 - C_{13}^4} }}} \right)^2}.
    	  \end{equation}	
    	For the bipartite pure states, the upper boundary for the blue solid curve is unchanging and always is  $D_{ij}^2 + C_{ij}^2 = 1$. Therefore, the green areas of BNBSs can be obtained when the value of concurrence is less than  $0.76$. In the same way, for the other reduced states ${\rho _{23}}(XXZ)$  and ${\rho _{12}}(XXZ)$ , the lower boundary for the black dashed curve in FIG. \ref{Fig.5} (b) can be disclosed as follow	
    		\begin{equation}\label{E54} 
    		D_{12}^2 = \frac{{13C_{12}^4 - 20C_{12}^2 + 8}}{{2{{\left( {C_{12}^2 - 2} \right)}^2}}}.
    		\end{equation}
    	Subsequently, the red areas of BNBSs can be revealed when the value of concurrence is less than  ${{\sqrt {\sqrt {41}  - 3} } \mathord{\left/{\vphantom {{\sqrt {\sqrt {41}  - 3} } 2}} \right.\kern-\nulldelimiterspace} 2}$. From Eqs. (\ref{E49}), (\ref{E50}) and (\ref{E52}), we still can unveil $D_{ij}^2 + C_{ij}^2 = {P_{ij}} \le 1$ in the renormalized spin-1/2 Heisenberg \textit{XXZ} model.

     \section{Conclusions} \label{sec5}
	    In this work, we have investigated the QRs of a tripartite \textit{W}-type state. It is indicated that the mutual relations among concurrence, the degree of coherence, purity and Bell nonlocality can be revealed. Surprisingly, exact quantitative relation among the degree of coherence, concurrence and purity is obtained in all two-qubit states derived from the tripartite \textit{W}-type states. Additionally, both the concurrence and Bell nonlocality increase at the expense of the degree of coherence. Then, we have derived exact lower and upper boundary conditions for Bell nonlocal and Bell local states, and identify the bipartite states revealed at these wishful boundaries. On this basis, we have illustrated the exact relation between Bell nonlocality and the degree of coherence for the considered bipartite entangled states in the tripartite \textit{W}-type states. Furthermore, we have investigated two specific scenarios: one scenario is a tripartite \textit{W}-type state under decoherence channel, we research how the PF channel impacts the mutual relations among these QRs.  It turned out that the proportion of the areas of BNBSs relative to BLBSs will decrease with the increase of decoherence strength. Besides, the concurrence decreases with the growing intensity of decoherence, whereas, the degree of coherence is not affected by decoherence, \textit{i.e.}, the degree of coherence is immune to the PF channel.  And the other one is a practical system for a renormalized spin-1/2 Heisenberg \textit{XXZ} model, some exact relation expressions among QRs and purity have also been obtained. These results will greatly rich one understand the intrinsic relation of QRs and make one better manipulate them to implement quantum information processing.
    	
     \section*{Acknowledgement} 
	    This work was supported by the National Science Foundation of China under Grant Nos. 11575001 and 61601002, Anhui Provincial Natural Science Foundation (Grant No. 1508085QF139) and Natural Science Foundation of Education Department of Anhui Province (Grant No. KJ2016SD49), and also the fund from CAS Key Laboratory of Quantum Information (Grant No. KQI201701).

    \end{document}